\documentclass[twocolumn, tighten, longauthor]{myaastex62}
\usepackage{amsmath}
\shorttitle{Hard limits on the number of bits per photon}
\shortauthors{Michael Hippke}
\begin{document}
\title{INTERSTELLAR COMMUNICATION. VIII. HARD LIMITS ON THE NUMBER OF BITS PER PHOTON}
\author[0000-0002-0794-6339]{Michael Hippke}
\affiliation{Sonneberg Observatory, Sternwartestr. 32, 96515 Sonneberg, Germany}
\email{michael@hippke.org}

\begin{abstract}
A photon can encode several bits of information based on an alphabet of its time of arrival, energy, and polarization. Heisenberg's uncertainty principle places a limit on measuring pairs of physical properties of a particle, limiting the maximal information efficiency to $<59$ bits per photon in practice, and $<171$ bits per photon at Planck energy, at a data rate of one photon per second.\\
\end{abstract}

\section{Introduction}
Photons are precious in interstellar communications due to the large distances involved \citep{2017arXiv170603795H}. Therefore, it is important to understand photon information efficiency (PIE): how many bits can be encoded into each photon? Using an alphabet based on the photon's dimensions (time of arrival, energy, and polarization), encoding schemes can be built. An introduction on how to calculate the number of bits per photon as a function of the alphabet size, noise, and losses was given in \citet{2017arXiv171205682H}. Now, we examine the ultimate limits based on Heisenberg's uncertainty principle, atomic surface smoothness limits, and Planck energy.

\section{Uncertainty principle}
Heisenberg's uncertainty principle states that ``the more precisely the position is determined, the less precisely the momentum is known in this instant, and vice versa'' \citep{Heisenberg1927}, so that
$\Delta E \, \Delta t \geq \hbar/2$ where $\Delta E$ is the standard deviation of the particle energy, $\Delta t$ is the time it takes the expectation value to change by one standard deviation, and $\hbar$ is the reduced Planck constant. A photon pulse with a temporal width $\Delta t$ can therefore not be monochromatic, but has a spectrum. Both are related through a Fourier transform, and it can be shown that \citep{Griffiths2004,Rulli2005}

\begin{equation}
\label{tmin}
\Delta t_{\rm min} \geq K \frac{\lambda_0^2}{\Delta \lambda\,c}
\end{equation}

where $\lambda_0$ is the central wavelength, $\Delta \lambda$ is the width of the spectrum (FWHM), $c$ is the speed of light and $K\approx0.441$ for a Gaussian pulse shape. For example, an optical laser pulse ($\lambda_0=500\,$nm) with a 10\,\% bandwidth ($\Delta \lambda=50\,$nm) has a minimum width of $\Delta t_{\rm min}\approx7.3\times10^{-15}\,$s, or $\approx7\,$fs.

\section{Minimum wavelength}
Decreasing $\lambda_0$ allows for shorter $\Delta t$. A practical limit comes from the fact that polished surfaces of diffraction-limited telescope mirrors or lenses require a surface smoothness smaller than the wavelength \citep{rayleigh,1935lett.book.....D}, precisely $< \lambda/4$ peak-to-valley as well as $< \lambda/14$ root-mean-square surface accuracy \citep{1894tdfa.book.....S,2008moed.book.....S}. The physical limit of materials is set by the atomic radius $\approx0.1\,$nm \citep{bohr1913}. Near the limit, X-ray focusing limits arise from scattering at the electrons of the atomic shell \citep{1948JOSA...38..766K,2004JaJAP..43.7311S} with limits of $\approx 0.03\,$nm \citep{yu1999surface}. This results in a theoretical physical limit for usable optics of $\lambda \gtrapprox 0.5$\,nm \citep{2017arXiv171105761H}. The limit can be surpassed by beam-forming with electromagnetic fields, e.g. using a free electron laser, however such methods are not energetically competitive \citep{2017arXiv171107962H}. In any case, the highest possible is the Planck energy, $E_{\rm P}=\sqrt{\hbar c^5/G}$ with $\lambda_0 \approx10^{-34}\,$m. Higher energy particles would directly collapse into a black hole.

\section{Maximum bandwidth}
Increasing bandwidth $\Delta \lambda$ allows for shorter $\Delta t$ (although PIE limits will be dominated by $\lambda_0$). Interstellar dispersion and scattering place upper limits on the usable wavelength, and therefore on the bandwidth.

Earth's ionosphere is opaque for frequencies $v<10$\,MHz ($\lambda>30$\,m), and the ionized interstellar medium in the galaxy absorbs radio signals $v<2$\,MHz ($\lambda>150$\,m) \citep{condon2016essential}. In addition, interstellar dispersion broadens pulses $\lambda>\mu$m \citep{1993ApJ...411..674T,2011AcAau..68..366S} to

\begin{equation}
\Delta t_{\rm disp} = 4\times10^{-15}\,{\rm DM}\,\lambda^2c^2\,\,\,{\rm (s)}
\end{equation}

where a typical value for the dispersion measure is $\rm{DM}=1$ over 100\,pc in the solar neighborhood, and $\rm{DM}=100$ over kpc towards the galactic center \citep{2017ApJ...835...29Y}. This limits pulse widths to $\Delta t_{\rm disp} > 10^{-8}\,$s at $\lambda=\,$m over pc distances to the nearest stars which is stricter than the uncertainty limit, $\Delta t_{\rm min} \gtrapprox 10^{-10}\,$s. Dedispersion is imperfect, because the true interference is unknown and changes with time \citep{alder2012radio}. Photon counting at radio frequencies is difficult near the the quantum limit due to the low energy per photon.

\begin{figure*}[ht]
\includegraphics[width=0.51\linewidth]{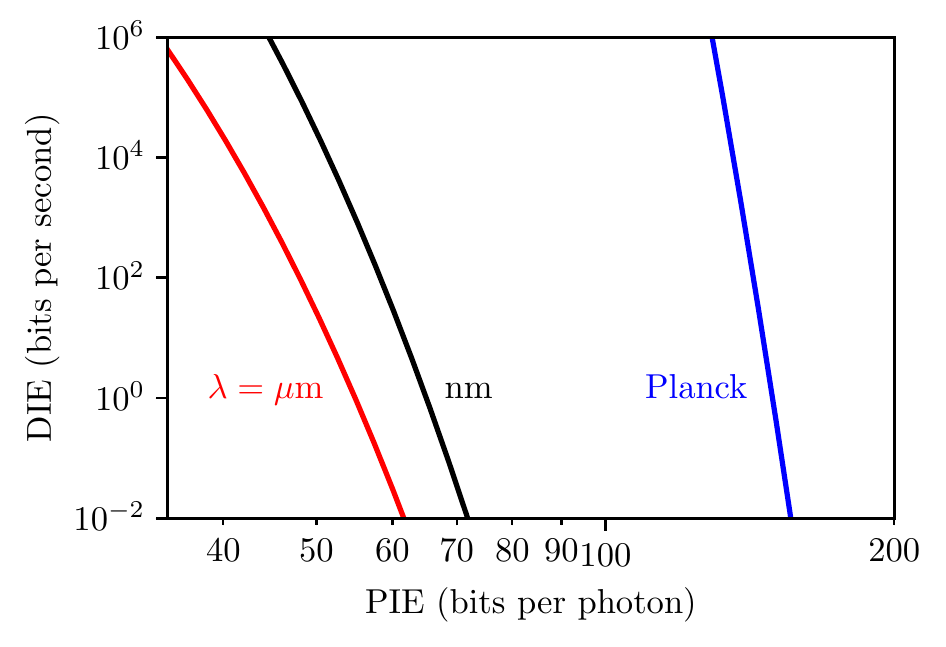}
\includegraphics[width=0.49\linewidth]{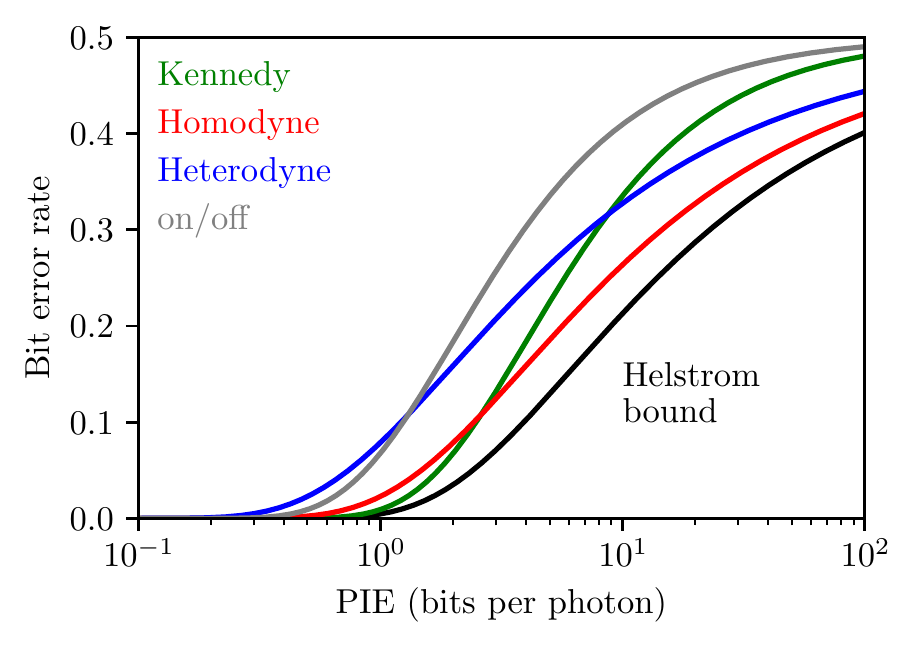}
\caption{\label{fig:pie_die}Left: Increasing PIE decreases dimensional information efficiency (DIE), and thus data rate. Values for $\Delta \lambda = \lambda_0$. Right: Bit error rates as a function of PIE for the Helstrom bound and different practical implementations.\\}
\end{figure*}

\pagebreak

\section{Information efficiency}
The ultimate noiseless quantum information efficiency is \citep{holevo1973bounds,2004PhRvL..92b7902G}

\begin{equation}
\label{cult}
{\rm PIE}=g(\eta M) \,\,\,{\rm (bits\,\,per\,\,photon)}
\end{equation}

where $M$ is the number of photons per mode, $\eta$ is the receiver efficiency and

\begin{equation}
\label{gx}{}
g(x)=(1+x) \log_2 (1+x)-x \log_2 x
\end{equation}

so that $g(x)$ is a function of $\eta \times M$. Without noise and for a perfect receiver ($\eta=1$) we can approximate

\begin{equation}
\label{eq_m-1}
{\rm PIE} \approx \log_2 (M^{-1})
\end{equation}

valid to within 5\,\% for $M^{-1}>10$. Heisenberg's uncertainty principle is an integral part of the Holevo bound, as the number of modes is finite for any finite amount of energy, time and space.

When we decrease $\lambda_0$ towards its minimum, the shortest $\Delta t_{\rm min}$ occurs for $\Delta \lambda = \lambda_0$, i.e. at 100\,\% bandwidth. When approximating $K=0.5$ in Equation~\ref{tmin} we get

\begin{equation}
\Delta t_{\rm min} \geq 0.5 \frac{\lambda_0^2}{\lambda_0\,c} = \frac{\lambda_0}{2c}.
\end{equation}

Each photon in a communication shall be transmitted (and received) within a finite amount of time, $t_{\rm dur}$. Within this time, the number of modes for this photon is then 

\begin{equation}
\label{eq_m}
M = \frac{\lambda_0}{2c\,t_{\rm dur}}.
\end{equation}

We can now insert Equation~\ref{eq_m} into Equation~\ref{eq_m-1} and get

\begin{equation}
\label{bitmax}
{\rm PIE} \lessapprox \log_2 \left(\frac{2c\,t_{\rm dur}}{ \lambda_0}\right)\,\,\,{\rm (bits\,\,per\,\,photon)}
\end{equation}

For $\lambda_0=1\,$nm and $t_{\rm dur}=1\,$s we find $C_{\rm ult}\approx59$ bits per photon, and 171 bits per photon at Planck energy. Including two alternative states in polarizations would double $M^{-1}$ and thus increase ${\rm PIE}$ by $\log_2(2)=1$ bit per photon.

\section{Trade-off between information efficiency and data rate}
Maximizing PIE comes at the cost of low dimensional information efficiency (DIE), measured in bits per mode, and thus data rate (in physical units: bits/sec/Hz). At the ultimate Holevo limit, PIE relates to DIE as \citep{2011arXiv1104.2643D} ${\rm PIE} = {\rm DIE} \times M^{-1}$. Figure~\ref{fig:pie_die} (left) shows the trade-off between PIE and DIE.

When one photon leverages $10^{18}$ time slots within one second, then only one photon can be received in this channel at the given PIE. If instead $10^{18}$ colors are used, the arrival of the photon can only be measured to within this one second. Thus, data rates are slow, but can be traded for higher DIE and thus higher data rates (but lower PIE). Conversely, one could further increase PIE by sending even fewer photons, e.g. one per year, which gives an additional $3\times10^7$ one-second slots, and thus increases PIE by $\approx23$ bits per photon.

\section{Bit errors}
The quantum nature of physical systems imposes an impossibility to discriminate perfectly between states. The lowest error probability allowed by quantum mechanics is the \citet{1976quantum} bound

\begin{equation}
P_{\rm Helstrom}=\frac{1}{2} \left( 1-\sqrt{1-e^{-4\, {\rm PIE^{-1}}}} \right).
\end{equation}

The performance of practical receiver implementations \citep{Wittmann2008,2011SPIE.8065E..0FG,cariolaro2015quantum} converge towards the bound in different regimes \citep{kennedy1973research,dolinaroptimum}. The resulting bit error rate as a function of PIE is shown in Figure~\ref{fig:pie_die} (right), and can be treated with software forward error correction at the expense of data rate \citep{moon2005error,huang2009software}. The extra overhead from using optimal correction codes is small (few percent).

\section{Discussion and conclusion}
In a practical interstellar communication, the distance between transmitter and receiver will be large causing significant diffractive beam widening, so that only a small fraction (typically $\ll 10^{-10}$) of the transmitted photons are collected by the receiver, ideally exactly one per pulse. As the source is unresolved in realistic interstellar communications, we neglected spatial encoding.

In practice, the maximum information efficiency is limited by non-zero thermal noise. For $\lambda=\mu$m and 100\,\% bandwidth, $\Delta t_{\rm min}\approx10^{-15}\,$s which encodes $C_{\rm ult}\approx49$ bits per photon. Current laboratory demonstrations achieve 19.3 bits per photon, or $\approx40\,$\,\% of the maximum, limited by dark noise in a $^3$He-cooled (350\,mK) superconducting detector \citep[section 3.1 in][]{2013SPIE.8610E..06F}.

\acknowledgments
\textit{Acknowledgments}
MH is thankful to Vittorio Giovannetti and David G. Messerschmitt for useful discussions.

\end{document}